\documentclass{article}
\usepackage{graphicx}
\usepackage{amsmath}
\usepackage{setspace}
\usepackage{amsfonts}
\usepackage{amssymb}
\providecommand{\U}[1]{\protect\rule{.1in}{.1in}}

\ifx\pdfoutput\relax\let\pdfoutput=\undefined\fi
\newcount\msipdfoutput
\ifx\pdfoutput\undefined\else
\ifcase\pdfoutput\else
\ifx\paperwidth\undefined\else
\ifdim\paperheight=0pt\relax\else\pdfpageheight\paperheight\fi
\ifdim\paperwidth=0pt\relax\else\pdfpagewidth\paperwidth\fi
\fi\fi\fi

\begin{document}

\title{\textbf{Self-Interacting Electromagnetic Fields and a Classical Discussion on
the Stability of the Electric Charge}}
\author{S.O. Vellozo$^{1,3}$ \thanks{E-mail: vellozo@cbpf.br}$\;$, Jos\'{e} A.
Helay\"{e}l-Neto$^{1,2}$ \thanks{E-mail: helayel@cbpf.br}$\;$,A.W. Smith$^{1}$
\thanks{E-mail: awsmith@cbpf.br}\\and L. P. G. De Assis$^{4,2,1}$ \footnote{E-mail: lpgassis@ufrrj.br}\\$^{1}$\textit{{\normalsize Centro Brasileiro de Pesquisas F\'{\i}sicas --
CBPF,}} \\\textit{{\normalsize Rua Dr.\ Xavier Sigaud 150, 22290-180, Rio de Janeiro,
RJ, Brasil}} \\$^{2}$\textit{{\normalsize Grupo de F\'{\i}sica Te\'{o}rica Jos\'e Leite
Lopes,}} \\\textit{{\normalsize P.O.\ Box 91933, 25685-970, Petr\'opolis, RJ, Brasil}} \\$^{3}$\textit{{\normalsize Centro Tecnologico do Ex\'{e}rcito -- CTEx }} \\\textit{{\normalsize Av. das Americas 28705, 230020-470, Rio de Janeiro, RJ,
Brasil}}\\$^{4}$\textit{{\normalsize Departamento de F\'{\i}sica, Universidade Federal
Rural do Rio de Janeiro }} \\\textit{{\normalsize BR 465-07, 23851-180, Serop\'{e}dica, Rio de Janeiro,
Brazil.}}}
\maketitle
\begin{abstract}
The present work proposes a discussion on the self-energy of charged particles
in the framework of nonlinear electrodynamics. We seek magnetically stable
solutions generated by purely electric charges whose electric and magnetic
fields are computed as solutions to the Born-Infeld equations. The approach
yields rich internal structures that can be described in terms of the physical
fields with explicit analytic solutions. This suggests that the anomalous field
probably originates from a magnetic excitation in the vacuum due to the
presence of the very intense electric field. In addition, the magnetic contribution has
been found to exert a negative pressure on the charge. This, in turn, balances the
electric repulsion, in such a way that the self-interaction of the field
appears as a simple and natural classical mechanism that is able to account
for the stability of the electron charge.
\end{abstract}

\pagebreak 

\section{INTRODUCTION}

\qquad By adopting a nonlinear approach to electrodynamics, in a previus work
we have found that an electric charge at rest generates a regular
magnetostatic field \cite{1}. The present work investigates how nonlinearity
can be used to reveal the presence of an intrinsic angular momentum and to
explain the mechanism that holds the electric charge together, ensuring its
stability. That finding is very interesting since a classical nonlinear
electrodynamics approach to describe the field interaction has naturally led
to the electronic spin. In addition a second major result is presented. The
calculations have shown that the field interacts with itself creating a
negative pressure in the charge that is strong enough to prevent it from
bursting. Such finding could be a solution to the historical problem of
electron stability. We also point aou that the non-linearity may be the key to
the understanding of a number of microscopic effects \cite{7}.

It still remains to be found if Maxwell's field equations are to be considered
approximations of a more general nonlinear electrodynamical theory. The most
physically unpleasant aspect of Coulomb's law is its singularity, that may
lead to unbounded field strengths inside charges and thus to an infinite
self-energy. Since extremely high electrostatic field strengths are to be
found in the vicinity of elementary charges, such regions cannot be accurately
described by linear electrodynamics and thus are likely to be associated to
departures from Coulomb`s law predictions.

This paper is outlined as follows. In the First Section, we briefly describe
the Born-Infeld (B-I) Electrodynamics\cite{2}\cite{3}\cite{4} magnetostatic
field solution and we set a correlation with experimental data for the
electron. In addition, it includes the calculation of the classical angular
momentum due to the intrinsic field and compares it with the value predicted
for the quantum spin of the electron. The Second Section describes the
calculation of the field pressure produced by the anomalous magnetostatic
field. Finally, the Third Section summarizes the main conclusions and our
Final Considerations.

\section{ANGULAR MOMENTUM FROM FIELD\newline SELF INTERACTION}

\qquad This section briefly describes the solution to the Born-Infeld
equations for a standstill electron as well as the calculation of its
intrinsic angular momentum.

According to the standard linear electrodynamics, the presence of a standstill
electric charged particle creates an electric field only regardless of its
strength. However, according to nonlinear electrodynamics, anomalous effects
may also occur due to self-interactions of the fields. The accurate
description of high field intensities in the vicinity of an electric charge
requires the use of a nonlinear approach. Born-Infeld Electrodynamics has been
found to be adequate to describe the fields of a charged particle under such
extreme condition.

Considering a static point-like electric charge at the origin, the solution to
the first Maxwell equation $\overrightarrow{\nabla}\cdot\overrightarrow
{D}=e\delta(\overrightarrow{x})$, with $e$ as the elementary charge, is the
electric induction $\overrightarrow{D}=\frac{e}{4\pi r^{2}}\widehat{r}$, that
is singular like the solution from a linear theory. If the magnetostatic
sector is allowed to become excited by intense electrostatic fields, the
Born-Infeld constitutive relation ensues. Under the assumption that the
induced magnetostatic field is always less than the maximum field strength
$b$, the Born-Infeld relationship simplifies, although leaving a residual
influence of the electric sector:%

\begin{equation}
\overrightarrow{H}=\frac{\overrightarrow{B}-\left(  \frac{\overrightarrow
{E}\cdot\overrightarrow{B}}{b^{2}}\right)  \overrightarrow{E}}{\sqrt
{1-\frac{\overrightarrow{E}^{2}-\overrightarrow{B}^{2}}{b^{2}}-\left(
\frac{\overrightarrow{E}\cdot\overrightarrow{B}}{b^{2}}\right)  ^{2}}}%
\overset{|\overrightarrow{B}|\ll b}{\longrightarrow}\frac{\overrightarrow{B}%
}{\sqrt{1-\frac{\overrightarrow{E}^{2}}{b^{2}}}}=\sqrt{1+\frac{\overrightarrow
{D}^{2}}{b^{2}}}\overrightarrow{B}, \label{1}%
\end{equation}%

\begin{equation}
\overrightarrow{B}=\frac{\overrightarrow{H}}{\sqrt{1+\frac{\overrightarrow
{D}^{2}}{b^{2}}}}. \label{2}%
\end{equation}
Also, the Maxwell equations $\overrightarrow{\nabla}\cdot\overrightarrow{B}=0$
and $\overrightarrow{\nabla}\times\overrightarrow{H}=\overrightarrow{0}$ will
complete the set of equations needed to describe the fields. Considering the
radial and polar components to be dependent on the radius and the polar angle,
the solution for the magnetic field polar component of $\overrightarrow{H}$
can be written as \cite{1}:%

\begin{equation}
H^{\theta}(r,\theta)=A\frac{f(x)}{r^{3}}\sin(\theta). \label{3}%
\end{equation}

The dimensionless function, $f$, called "form function", describes the
transition from the linear to the nonlinear regime and its asymptotic limit is constant:%

\begin{equation}
f(x)=\frac{x^{2}}{0.3234}\left\{  \sqrt{x}P_{1/4}^{1/4}\left(  \sqrt{1+x^{4}%
}\right)  -\kappa x\right\}  \overset{r\rightarrow\infty}{\longrightarrow}1.
\label{4}%
\end{equation}

The function $P_{1/4}^{1/4}(z)$ is the associated Legendre function of first
kind and $\kappa(\approx0.82217)$ is the constant that ensures $H^{\theta}$ to
vanish when $r$ approaches infinity. Far away from the electric charge, $r\gg
r_{o}$, $H^{\theta}$ becomes a genuine magnetic dipole moment field given by:%

\begin{equation}
H^{\theta}\overset{r\rightarrow\infty}{\longrightarrow}\frac{A}{r^{3}}%
\sin(\theta) \label{5}%
\end{equation}

Dimensionally, the constant $A$ has units of a magnetic dipole so that
equation (\ref{5}) describes the macroscopic view of an intrinsic magnetic
dipole moment for the charge considered. Thus, in order to obtain a more
realistic solution, the constant $A$ can be assumed to correspond to the
intrinsic electron magnetic dipole moment, that is very close to the value of
Bohr's Magneton, $\mu_{Bohr}=9.27\times10^{-24}JT^{-1}$, in the MKS System.
Such assumption is needed so that we can bring input to our proposal.

Using the constitutive relation (\ref{2}), with $B^{j}\ll b$, the magnetic
induction can be set equal to \cite{1}:%

\begin{equation}
B^{\theta}(r,\theta)=\frac{H^{\theta}(r,\theta)}{\sqrt{1+\frac{\overrightarrow
{D}^{2}}{b^{2}}}}=\frac{\mu_{Bohr}}{r^{3}}\frac{x^{2}f(x)}{\sqrt{1+x^{4}}}%
\sin(\theta). \label{6}%
\end{equation}

Once the field structure has been determined, the angular momentum associated
with the stationary electric charge can be calculated. It is produced by the
interaction between the electric and the magnetic fields, which generates an
intrinsic angular momentum given by:%

\begin{equation}
\overrightarrow{L\text{ }}=\int\overrightarrow{x}\times\left(  \overrightarrow
{D}\times\overrightarrow{B}\right)  d^{3}\overrightarrow{x}. \label{7}%
\end{equation}

It must be highlighted that both fields inside integral (\ref{7}) are
generated by the point-like electric charge. Thus the magnetostatic induction,
$\overrightarrow{B}$, is to be regarded as a product of the nonlinearity only.
The simple assumption that the intense electric field caused by the electric
charge at rest can excite the magnetostatic sector and yield an intrinsic
field angular momentum is completely ruled out in any linear approach.

Back to equation (\ref{7}), only the axial component will be present due to
symmetry considerations and allowing $\overrightarrow{L}$ to be projected on
the axial dipole axis, the integral becomes:%

\[
L_{z}=\int\left(  r\right)  \left(  \frac{e}{4\pi r^{2}}\right)  \left[
\mu_{o}\left(  1+\frac{\overrightarrow{D}^{2}}{b^{2}}\right)  ^{-1/2}\frac
{\mu_{Bohr}}{r^{3}}f(x)\sin(\theta)\right]
\]%

\begin{equation}
\times\sin(\theta)r^{2}\sin(\theta)d\theta d\varphi dr. \label{7a}%
\end{equation}

This integral can be evaluated and written in compact form for Born-Infeld
parameters and natural constants. Defining%

\[
\gamma=\int\frac{f(x)dx}{\sqrt{1+x^{4}}}=1.18,
\]
and the B-I radius, recalculated by Born and Schr\"{o}dinger \cite{6} as:%

\[
r_{o}\simeq2.618\times10^{-14}m,
\]
then the axial component can be written as:%

\begin{equation}
L_{z}=\frac{2}{3}\left(  \frac{\gamma}{r_{o}}\right)  \left(  e\mu_{o}%
\mu_{Bohr}\right)  \simeq0.556\times10^{-34}Js. \label{9}%
\end{equation}

This value for the spin of the electron, obtained on purely classical grounds,
departs about 5\% only in comparison with the prediction from Quantum
Mechanics, $\hslash/2$. It is driven by the nonlinearity. Inside the first
parenthesis are the B-I parameters while enclosed in the second one are
natural constants. No mechanical rotation or other kind of translation has
been considered in order to generate $\overrightarrow{L}$, so that the angular
momentum appears naturally, as a remarkable consequence of the
self-interaction of the fields. The importance of this result lies not only in
its numerical value, but in how the charge produces its intrinsic angular
momentum, interpreted here as the spin of the charged particle.

The net result of this section is that the interaction between the electric
sector and the magnetic sector generates a spin.

\section{THE ELECTRIC CHARGE STABILITY}

\qquad This section tackles the delicate issue \ concerning the stability of
the electric charge. We present here the details of our claim: the field
self-interaction is responsible for the electric charge stability.

The dynamical properties of the electromagnetic field are described by the
energy-momentum tensor, $T^{\mu\nu}$. Among its components, $T^{rr}$ is of
particular interest because it expresses the radial force per unit area. In
terms of Born-Infeld Lagrangian, $\mathcal{L}_{BI}$, $T^{ij}$ \cite{5} is
written as:%

\begin{equation}
T^{ij}=-E^{i}D^{j}-H^{i}B^{j}+\delta^{ij}\left\{  \mathcal{L}_{BI}%
+\overrightarrow{H}\cdot\overrightarrow{B}\right\}  . \label{10}%
\end{equation}

Only the $T^{rr}$-component is of interest, since the others will not
contribute to the radial pressure:%

\begin{equation}
T^{rr}=-E^{r}D^{r}+\mathcal{L}_{BI}+H^{\theta}B^{\theta}. \label{11}%
\end{equation}

Integrating its projection over axial dipole axis, it becomes, in MKS System:%

\begin{equation}%
{\displaystyle\int\limits}
_{\substack{hemisphere\\surface}}\left(  dS\widehat{r}\right)  T^{rr}\left(
\widehat{r}\cdot\widehat{z}\right)  =\epsilon_{o}b^{2}\pi r_{o}^{2}%
P(x),\label{12}%
\end{equation}
where%

\begin{align}
P(x) &  =-\frac{1}{\sqrt{1+x^{4}}}+\left(  1-\frac{x^{2}}{\sqrt{1+x^{4}}%
}\right)  x^{2}+\label{13}\\
&  +\frac{1}{2}\left(  \frac{4\pi\mu_{Bohr}}{ecr_{o}}\right)  ^{2}\frac
{f^{2}(x)}{x^{2}\sqrt{1+x^{4}}}.\nonumber
\end{align}

The constants, $\epsilon_{o}$ and $c$, in (\ref{12}), are the vacuum electric
permittivity and speed of light, respectively. The term $\epsilon_{o}b^{2}$ is
the characteristic field pressure and its module is about $10^{25}N/m^{2}$.
This is a very high pressure. Considering the hemisphere area, $2\pi r_{o}%
^{2}$, where the integration will be performed, the intensity of that
particular force is in the order of $10^{-2}N$. In\ addition, the function
$P(x)$ expresses the competition between the outward electrical repulsion and
the inward magnetostatic pressures acting on the spherical surphace. The last
term, inside $P(x)$, is due to the self interaction of the magnetostatic
field. In its absence, the pressure becomes purely repulsive, regardless the
sign of the electrical charge. In contrast, its presence promotes a drastic
change. All calculations were carefully performed in MKS units system.
The balance between forces is depicted in $Figure\ 1$ for a hemisphere. It clearly
shows the change in sign of the net pressure as well as the drastic changes in
its magnitude.

%

\begin{center}
\begin{figure}[h]
\vspace{0.3cm}
{\par\centering
\resizebox*{1.00\textwidth}{!}{\rotatebox{-90}{\includegraphics{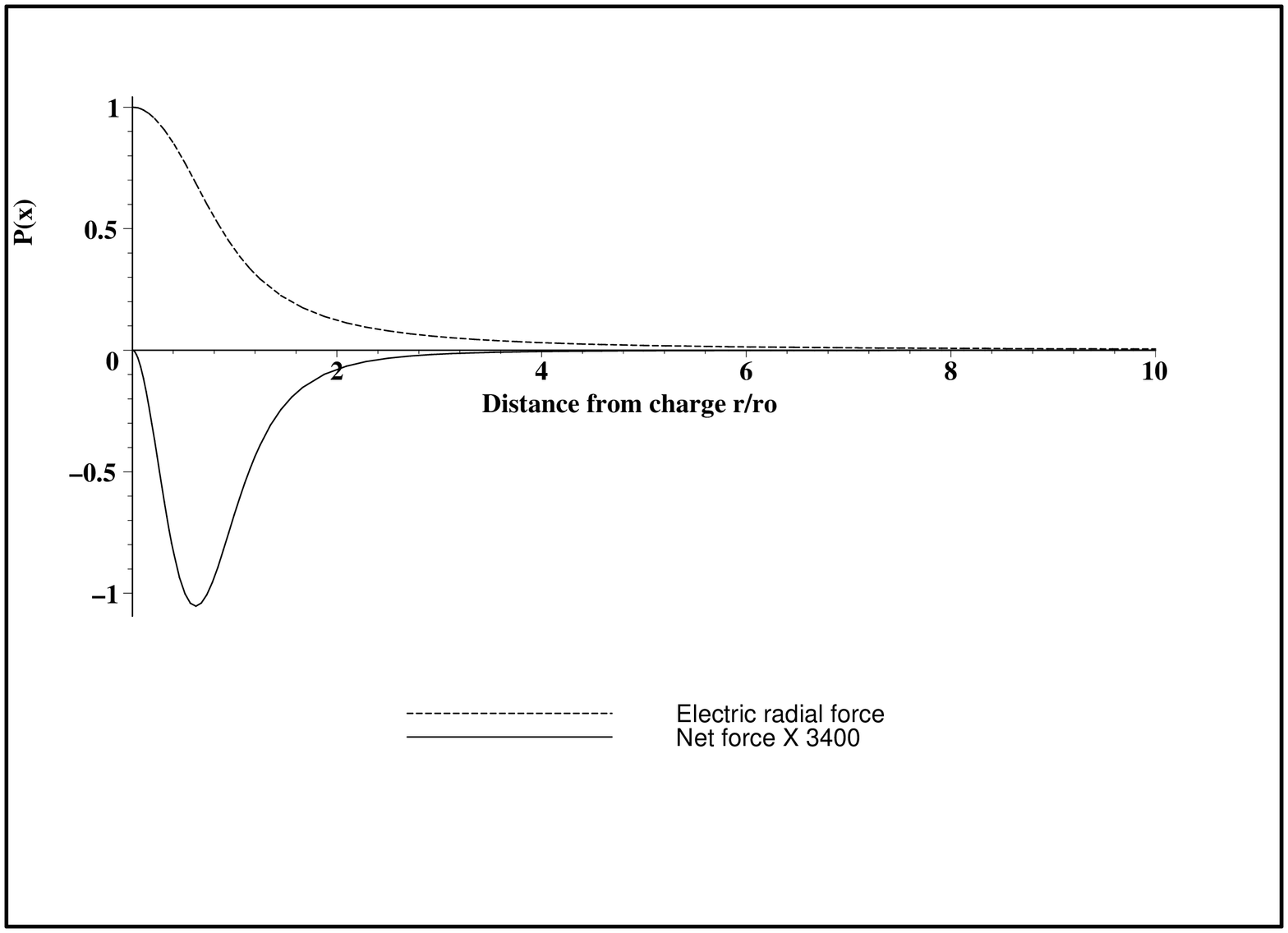}}}
\par}
{\par\ \par}
{\par\centering Figure 1\par}
\vspace{1cm}
\end{figure}
\end{center}

\bigskip
Closing this section the following list summarizes the major nonlinear\\
electrodynamics effects and their causes.

\begin{center}%
\begin{tabular}
[c]{|lll|}\hline
\ \ \ \ \ \ \ \ \ \ \ \ \ \ \ \ \ \ \ \ Effect &  &
\ \ \ \ \ \ \ \ \ \ \ \ \ \ \ \ \ \ \ \ \ Cause\\
Anomalous Magnetostatic Fields & $\Rightarrow$ & Extremely High Electrostatic
Field\\
Intrinsic Angular Momentum (Spin) & $\Rightarrow$ & Electric and Magnetic
Field Interaction\\
Charge Stability & $\Rightarrow$ & Magnetostatic field
Self-Interaction\\\hline
\end{tabular}
\end{center}

\pagebreak

\section{FINAL CONSIDERATIONS}

\qquad The results presented illustrate how accurately non-linearity can
represent physical phenomena. However, in spite of the apparent
self-consistency of this work, it must be stressed that it would be premature
to claim that it actually presents a legitimate description of Nature. The
fact that the intrinsic angular momentum of the electron, its spin, could be
predicted with a deviation of about 5\% only, suggests that predictions for
the net pressure are consistent. In other words, the stability of the
electronic charge may be described in terms of the self-interaction of the
magnetostatic fields. In the present paper, it has been proven that a
nonlinear self-interaction mechanism can explain the stability of the charge
distribution. Even if it yields some considerable difference with respect to
mesurable value, we believe that, qualitatively, it is relevant to understand
(classically) how self-interaction and stability are related. This is a lesson
we can implement in the framework of Yang-Mills theories.

A further step may be taken at this point. Expression (\ref{13}) can be
rewritten as a function of electronic spin $\overrightarrow{L}$and thus be
differently interpreted. Setting $\frac{\mu_{Bohr}}{r_{o}}=\frac{3L_{z}%
}{2\gamma e\mu_{o}}$ from (\ref{9}), and inserting it in (\ref{13}), yields a
connection between the pressure and the spin. Difining $u(x)$ and $v(x)$ as:%

\[
u(x)=-\frac{1}{\sqrt{1+x^{4}}}+\left(  1-\frac{x^{2}}{\sqrt{1+x^{4}}}\right)
x^{2},
\]

\[
v(x)=\left(  \frac{6\pi}{e^{2}c\mu_{o}}\right)  ^{2}\frac{f^{2}(x)}{x^{2}%
\sqrt{1+x^{4}}},
\]
leads to%

\begin{equation}
P(x)=u(x)+v(x)L_{z}^{2}. \label{14}%
\end{equation}

It can be easily seen that the term $v(x)L_{z}^{2}$ guarantees the stability
of the electric charge. It must then be concluded that the presence of spin is
necessary to ensure the integrity of the elementary charged particle and that
a spinless (truly) elementary charged particle is not expected to exist. This
result is in perfect agreement with the fact that no spinless charged (trully elementary)
particle has been discovered so far. However, the Minimal Supersymmetric
Standard Model (MSSM) predicts the existence of two charged spinless Higgs
bosons, in disagreement with the approach proposed here. On the other hand,
such particles have not been detected yet and still remain as a theoretical
possibility. There is also the possibility that, once they are found (at LHC,
for instance), they turn out to be composite structures and not genuinely
elementary particles.

\bigskip \noindent{\bf Acknowledgments}\\
S.O.V. wish to thanks to CBPF for Academic support and CTEx by the financial help. L.P.G.A is
grateful to FAPERJ-Rio de Janeiro for his post-doctoral fellowship. J.A.H.-N. expresses his gratitude to CNPq for financial help.

\pagebreak

\end{document}